\documentclass[12pt,preprint]{aastex}

\usepackage{apjfonts}

\shorttitle{The complete spectrum of the neutron star X-ray binary 4U 0614+091}
\shortauthors{Migliari et al.}

\begin{document}

\title{The complete spectrum of the neutron star X-ray binary 4U~0614+091} 
\author{S. Migliari\altaffilmark{1,2}, J.A. Tomsick\altaffilmark{3}, J.C.A. Miller-Jones\altaffilmark{4}, S. Heinz\altaffilmark{5}, R.I. Hynes\altaffilmark{6}, R.P. Fender\altaffilmark{7}, E. Gallo\altaffilmark{8}, P.G. Jonker\altaffilmark{9,10}, T.J. Maccarone\altaffilmark{7}}





\altaffiltext{1}{European Space Astronomy Centre, Apartado/P.O. Box 78, Villanueva de la Canada, E-28691 Madrid, Spain}
\altaffiltext{2}{Center for Astrophysics and Space Sciences, University of California San  Diego, 9500 Gilman Dr., La Jolla, CA 92093-0424, USA}
\altaffiltext{3}{Space Sciences Laboratory, 7 Gauss Way, University of California, Berkeley, CA 94720-7450, USA}
\altaffiltext{4}{National Radio Astronomy Observatory, 520 Edgemont Road, Charlottesville, VA 22903, USA}
\altaffiltext{5}{Department of Astronomy, University of Wisconsin-Madison, 6508 Sterling Hall, 475 North Charter Street, Madison, WI 53593, USA}
\altaffiltext{6}{Department of Physics and Astronomy, Louisiana State University, Baton Rouge, Louisiana 70803, USA}
\altaffiltext{7}{School of Physics and Astronomy, University of Southampton, Hampshire SO17 1BJ, United Kingdom}
\altaffiltext{8}{MIT Kavli Institute for Astrophysics and Space Research, 70 Vassar St., Cambridge, MA 02139, USA}
\altaffiltext{9}{SRON, Netherlands Institute for Space Research, 3584 CA Utrecht, the Netherlands}
\altaffiltext{10}{Harvard--Smithsonian Center for Astrophysics, 60 Garden Street, Cambridge, MA~02138, U.S.A.}
\begin{abstract}

We observed the neutron star (NS) ultra-compact X-ray binary 4U~0614+091 quasi-simultaneously in the radio band (VLA), mid-IR/IR (Spitzer/MIPS and IRAC), near-IR/optical (SMARTS), optical-UV (Swift/UVOT), soft and hard X-rays (Swift/XRT and RXTE). The source was steadily in its `hard state'. We detected the source in the whole range,  for the first time in the radio band at 4.86 and 8.46~GHz and in the mid-IR at 24 $\mu$m, up to 100 keV.  The optically thick synchrotron spectrum of the jet is consistent with being flat from the radio to the mid-IR band. The flat jet spectrum breaks in the range $\sim(1-4)\times10^{13}$~Hz to an optically-thin power-law synchrotron spectrum with spectral index $\sim-0.5$.  These observations allow us to estimate a lower limit on the jet radiative power of $\sim3\times10^{32}$~erg s$^{-1}$ and a total jet power L$_{J}\sim10^{34} \mu_{0.05}^{-1}$ E$_{c}^{0.53}$~erg/s  (where E$_{c}$ is the high-energy cutoff of the synchrotron spectrum in eV and $\mu_{0.05}$ is the radiative efficiency in units of 0.05). The contemporaneous detection of the optically thin part of the compact jet and  the X-ray tail above 30~keV allows us to assess the contribution of the jet to the hard X-ray tail by synchrotron self-Compton (SSC) processes. We conclude that, for realistic jet size, boosting, viewing angle and energy partition, the SSC emission alone, from the post-shock, accelerated, non-thermal population in the jet, is not a viable mechanism to explain the observed hard X-ray tail of the neutron star 4U~0614+091. 

\end{abstract}

\keywords{radio continuum: stars - infrared: general - X-rays: binaries - accretion, accretion disks - ISM: jets and outflows - stars: individual (4U~0614+091)}

\section{Introduction}

Relativistic jets have been observed in a number of X-ray binaries (XRBs) containing  black holes (BHs) and neutron stars (NSs; see Fender 2006 for a review). Recently, K\"ording et al. (2008) found evidence for a transient jet in a binary system containing a white dwarf, confirming that the existence of jets in such systems does not depend on the nature of the compact object. The most studied and best known jet sources are the BH systems, because they  are typically much brighter in the radio band (Migliari \& Fender 2006; K\"ording et al. 2007). The radio band is the observing window where the jet emission routinely dominates over the other radiative components in XRBs (companion star, disk). 
Low-luminosity BHs - less than  a few \% of the Eddington limit - are characterized by an optically thick, flat or slightly inverted, synchrotron radio spectrum ($\alpha\gtrsim0$, where $F_{\nu}\propto \nu^{\alpha}$ and $F_{\nu}$ is the flux density at the frequency $\nu$; Fender 2001). This optically thick synchrotron emission is interpreted as a spectral signature of conical continuously replenished, self-absorbed compact jets (developed to model steady jets in AGN: Blandford \& K\"onigl 1979; see also Hjellming \& Johnston 1988; Falcke \& Biermann 1995; Kaiser 2006). This interpretation has been confirmed by radio imaging of the milli-arcsec scale jets, for two Galactic BH XRBs: Cyg X-1 (Stirling et al. 2001) and GRS~1915+105  (Dhawan, Mirabel \& Rodr\'iguez 2000).
At shorter wavelengths, compact jet models predict a `break' from an optically thick to an optically thin  synchrotron spectrum ($\alpha\sim-0.6$). This optically thin spectrum represents the emission from the part of the jet that is closer to the base which is transparent to its emitting synchrotron radiation. This break of the jet spectrum has been inferred with spectral fitting to be in the IR/optical range (Nowak et al. 2005), and actually detected in two BHs, GX~339-4 (Corbel \& Fender 2002) and XTE J1118+480 (Hynes et al. 2006).
The accretion state where the compact jet is active corresponds to the `hard state' as defined in McClintock \& Remillard (2006) or Homan \& Belloni (2005).  In such a state, the X-ray band is dominated by a non-thermal power-law emission, which can be detectable up to hundreds of keV. Most of the BHs show a cut-off around 100-200 keV, but an extra power-law extending up the MeV has also been observed in Cyg~X-1(e.g. McConnel et al. 1994). Different scenarios have been proposed to describe the origin of this hard X-ray tail, with the main processes being inverse Compton scattering from a population of thermal or hybrid thermal/non-thermal hot electrons in a `corona' whose geometry is still an area of debate (e.g. Narayan et al. 1997; Esin et al. 1996; Poutanen \& Coppi 1998; Merloni 2003; Poutanen \& Vurm 2009; see also Titarchuk et al. 1996 for the bulk motion Comptonization proposal), synchrotron self-Compton (SSC) emission from the base of a jet, or a combination of the two (e.g. Markoff, Nowak \& Wilms 2005; Migliari et al. 2007). 

The jet phenomenology observed in BHs can be found also in NS X-ray binaries. The radio detection of low-luminosity NS XRBs and  brightness temperature arguments strongly indicate that also NSs in their `hard state' produce jets (Migliari et al. 2003). In particular, the detection of the infra-red (IR) optically-thin break in the synchrotron spectrum of 4U~0614+091 brought the evidence that a jet of the same kind observed in AGN and Galactic BHs, modeled as a compact jet {\it a la} Blandford \& K\"onigl (1979) can be formed in a NS system (Migliari et al. 2006). From a direct comparison of the ratio of radiative jet power to X-ray bolometric power, L$_{radio}$/L$_{Xbol}$, in 4U~0614+091 and in BHs, we found that in BHs  L$_{radio}$/L$_{Xbol}$ is about two orders of magnitude higher. This difference might be explained by a difference in the role of the jet as a power output channel in the two types of systems: BHs are jet-dominated (Fender et al. 2004) while NSs never enter a jet dominated regime (Migliari \& Fender 2006).

NSs, like BHs, also show non-thermal hard X-ray tails up to hundreds of keV, typically with cutoff at lower energies than BHs, at around 50~keV, which are well modeled by thermal Comptonization in a hot corona (e.g. Barret et al . 2000; Medvedev \& Narayan 2001; Maccarone \& Coppi 2003; Done \& Gierli\'nski 2003; Paizis et al. 2006 and references therein for bright Z-type NSs; e.g. Fiocchi et al. 2008 for 4U~0614+091).  The contemporaneous detection of the hard tail and the optically-thin part of the jet would give crucial hints on the possible role of the jet in the hard non-thermal X-ray emission of NSs. 

4U~0614+091 is an ultra-compact NS XRB (Bradt et al. 1992; Juett et al. 2001), with a possible period of 51~min (Shahbaz et al 2008a), and is the closest to us at a distance of 3.2~kpc (the distance was recently estimated, with an error of $\sim15\%$, thanks to the observation of an X-ray burst showing a photospheric radius expansion: Kuulkers et al. 2009, Kuulkers et al 2003). Given the ultra-compact nature of the system, the small accretion disk and the faint white dwarf or He-rich partially degenerated companion do not emit significantly in the mid-IR band, which is dominated by the optically thin emission of the jet. 4U~0614+091 is also a hard X-ray source, showing non-thermal emission above 100~keV (e.g. Fiocchi et al. 2008). The source is about 90\% of its time in a steady `hard state', as defined for atoll-type sources (e.g. van Straaten et al. 2000). Making a parallel with BHs, we expect the compact jet to be active in this state.

\section{Data Analysis}

We have observed the NS XRB 4U~0614+091 simultaneously in the radio band with the Very Large Array (VLA), in mid-infrared and infrared (IR) with the Spitzer Space Telescope, in near-IR/optical with the ground-based Small and Moderate Aperture Research Telescope System (SMARTS), in the optical and soft X-rays with Swift and in soft and hard X-rays with the Rossi-X-ray Timing Explorer (RXTE), covering a range from 4.86~GHz to 200~keV.

\subsection{VLA}\label{data_vla}

4U\,0614+091 was observed with the VLA in its relatively compact C-configuration on November 1--2, 2006. The observation at 8.46 GHz started at 09:26:45 UT with an exposure time of 290.4 min and the observation at 4.86 GHz started at 05:07:58 UT with an exposure time of 290.7 min.  3C\,48 was used as the primary calibrator, setting the flux density scale according to coefficients derived at the VLA in 1999.2 by staff at National Radio Astronomy Observatory (NRAO).  The secondary calibrator was J\,0613+1306, 4.0$^{\circ}$ away from the target source.  Six of the twenty-six telescopes in the array were retrofitted EVLA antennas, necessitating the determination of baseline-based gain solutions to prevent closure errors on EVLA-VLA baselines due to mismatched bandpass shapes.  The observing frequencies were 8.46 and 4.86\,GHz, and the observing bandwidth was 50\,MHz in both cases.  Data were reduced according to standard procedures in the 31Dec08 version of {\textsc AIPS} (Greisen 2003) using a clean component model to determine the gain solutions for the primary calibrator.  Since the source was so faint, no self-calibration was performed.  The source flux density and position were measured after deconvolution by fitting the source as a point source and as an elliptical Gaussian in the image plane.

\subsection{Spitzer: MIPS and IRAC}\label{data_spitzer}

We observed 4U~0614+091 with Spitzer/IRAC on October 30 2006 starting at 17:41:39 UT  for 384 s and with Spitzer/MIPS on November 3, 2006 starting at  06:57:43 UT  for 3371 s.  
We have processed the $24~\mu$m MIPS and the 3.6, 4.5, 5.8 and 8~$\mu$m IRAC Basic Calibrated Data using the software {\tt mopex} (Makovoz \& Marleau 2005), following the procedure as in the manuals available on line\footnote{Available at the {\it Spitzer} Science Center (SSC) web-page http://ssc.spitzer.caltech.edu}. We created a mosaic from the 280 and 10 frames per band obtained in the MIPS and IRAC observations, respectively, 
and we extracted the source flux using aperture photometry.  
We have corrected the IRAC flux densities for interstellar extinction using $A_{\rm v}=2$ [derived from the equivalent hydrogen column density values in Juett, Psaltis \& Chakrabarty (2001) and using $N_{\rm H}=A_{\rm v}\times0.179\times10^{22}$ in Predehl \& Schmitt (1995)], and following the standard optical-to-IR interstellar extinction law with 0.55~$\mu$m as a reference frequency for the V band (Rieke \& Lebofsky 1985; Cardelli, Clayton \& Mathis 1989).
These de-reddening values calculated from the X-ray spectra have to be considered upper limits (Juett et al. 2001). However, studies of the optical spectra of the source indicate that the reddening of the object should be close to the maximum in the Galaxy in its direction (see Nelemans et al. 2004). Therefore, we use these values (also adopted in Nelemans et al. 2004) as a fair approximation also for SMARTS and Swift/UVOT flux corrections, with an uncertainty of $\sim30\%$ (Bohlin et al. 1978). For Spitzer, the corrections for interstellar extinction are small, i.e. $\sim10\%$ for the flux density at $3.6 \mu$m and less than 5\% for the flux densities in the other three IRAC bands. No extinction correction was made at $24~\mu$m. We add, quadratically to the rms, a 5\% and 10\% systematic error on the estimate of the flux densities in the IRAC and MIPS observations, respectively, to take into account the uncertainties on the photometric calibration (see Reach et al. 2005).

\subsection{SMARTS 1.3m: Andicam}\label{data_smarts}

4U~0614+091 was observed on October 30, 2006 at 05:31 UT  for 660s using the Andicam dual-channel optical/IR camera on the SMARTS 1.3\,m telescope at Cerro-Tololo, Chile.  Optical photometry was obtained in the $V$ and $I$ bands, and infrared in the $J$ band.  The data were reduced in IRAF\footnote{IRAF is distributed by the National Optical Astronomy Observatories, which are operated by the Association of Universities for Research in Astronomy, Inc., under cooperative agreement with the National Science Foundation.} using standard procedures.  In the optical, differential photometry was performed relative to a bright comparison star in the field, and this star was then calibrated on several nights in 2007 January relative to multiple stars in the standard field Ru~149 (Landolt 1992).  IR data were obtained in a 5-point dither pattern, and combined before measuring the brightness of the target relative to a nearby star in the field from the 2MASS catalog. We have applied de-reddening corrections as described in \S~\ref{data_spitzer}.

\subsection{Swift: UVOT and XRT}\label{data_swift}

We analyzed the {\em Swift} X-ray Telescope (XRT) and UV-Optical Telescope (UVOT) data from the observation of 4U~0614+091 that started on 2006 October 30, at 17:36 UT for 3256 s (ObsID 00030812001).  We used the Windowed Timing mode for the XRT and obtained a total XRT exposure time of 3,256 s.  We reduced the XRT data in 2008 May using the HEASOFT v6.4.1 software.  We used the HEASOFT routine {\ttfamily xselect} to produce source and background spectra. For the source spectrum, we included photons from the cleaned event list that are within $47^{\prime\prime}$ of
the source centroid, and we used a background region that is the same size but is offset by $2^{\prime}.8$.  After background subtraction, we measured a count rate of 44 counts~s$^{-1}$ for the source.  We generated a response using the calibration files that were current in 2008 May. For the UVOT data, we started by using the HEASOFT routine {\ttfamily uvotimsum} to sum the images for the different filters used.  Then, we defined source and background regions and used {\ttfamily uvot2pha} to produce the spectra and response files, from which we obtained the flux measurements given in Table~1 for the V, B, U, UVW1, and UVM2 filters. We have applied de-reddening corrections as described in \S~\ref{data_spitzer}. Note that, contrary to the IR band, the extinction correction in the UV spectrum are large, up to 26\%. Therefore, the uncertainty on the de-reddening value can be significant.

\subsection{RXTE}\label{data_rxte}

For the energy spectral analysis, we combined the data taken on  the 30 October, 2008, i.e. the observations  92411-01-06-07 and 92411-01-06-00 for a total exposure time of 4160 s. We have used PCA {\tt Standard2} of all the PCUs available,  and HEXTE  {\tt Standard Mode} cluster B data. For the PCA data, we have  subtracted the background estimated using {\tt pcabackest} v.3.0,  produced the detector response matrix with {\tt pcarsp} v.10.1, and  analyzed the energy spectra in the range 3--20~keV. A systematic error of 0.5\% was added to account for uncertainties in the  calibration. For the HEXTE data, we corrected for deadtime, subtracted  the background, extracted the response matrix using HEASOFT v.6.4.1.  We have analyzed the HEXTE spectra between 20 and $200$~keV. 

\rotate{
\begin{table*}[htdp]
\footnotesize{

\begin{center}
\begin{tabular}{cccccc}
\hline
\hline
Telescope & Date & Start Time (UT) & Exposure time (sec) & Energy band & Flux\\
\hline
VLA & 2006 Nov 1-2 & 09:26:45 & 17,424  & 4.86\,GHz &  $0.281\pm0.014$ mJy \\
       & 2006 Nov 1-2 & 05:07:58 & 17,442  & 8.46\,GHz &  $0.276\pm0.010$ mJy \\
\hline
Spitzer/MIPS & 2006 Nov 03 & 06:57:43 & 3371 & 24\,$\mu$m & $0.35\pm0.06$ mJy \\
\hline
Spitzer/IRAC & 2006 Oct 30 & 17:41:39 & 384  &  8\,$\mu$m & $0.26\pm0.02$ mJy \\
                     & 2006 Oct 30 &  & &  5.8\,$\mu$m & $0.20\pm0.02$ mJy \\
                     & 2006 Oct 30 &  & &  4.5\,$\mu$m & $0.20\pm0.02$ mJy \\
                     & 2006 Oct 30 &  & &  3.6\,$\mu$m & $0.17\pm0.01$ mJy \\
\hline
SMARTS & 2006 October 30 & 05:31 & 660   & V (0.55 $\mu$m)& $0.956\pm0.002$  \\
               & 2006 October 30 &  & & I (0.81 $\mu$m)&  $0.397\pm0.001$  \\
               & 2006 October 30 & & & J (1.25 $\mu$m)&  $0.244\pm0.001$  \\

\hline
Swift/UVOT
& 2006 Oct 30 & 17:36 & 3,256  & V (0.54 $\mu$m) & $1.27\pm0.27$ mJy \\
                    & 2006 Oct 30 & & &B (0.44 $\mu$m) & $1.06\pm0.15$ mJy \\
                    & 2006 Oct 30 & & &U (0.35 $\mu$m) & $1.04\pm0.08$ mJy \\
                    & 2006 Oct 30 & & & UVW1 (0.26 $\mu$m) & $0.85\pm0.07$ mJy \\
                   & 2006 Oct 30 &  & & UVM2 (0.22 $\mu$m) & $0.73\pm0.09$ mJy \\

\hline
Swift/XRT    & 2006 Oct 30 & 17:36 & 3,256 s & 0.6-10\,keV & $2.6\pm0.5$\,($\times10^{-9}$ erg s$^{-1}$ cm$^{-2}$)\\

\hline
RXTE/PCA        & 2006 Oct 30 & averaged & 4160  &  3-20\,keV & $9.7\pm0.1$\,($\times10^{-10}$ erg s$^{-1}$ cm$^{-2}$)  \\
RXTE/HEXTE    & 2006 Oct 30 & &  & 20-200\,keV$^{a}$ & $3.3\pm0.1$\,($\times10^{-10}$ erg s$^{-1}$ cm$^{-2}$)  \\

\end{tabular} \caption{ Log of the quasi-simultaneous multiwavelength observations of 4U~0614+091: telescope, date, start time, duration, wavelength and flux or flux density. IR flux densities are not de-reddened. {\bf Notes.} 
$^{a}$ the source is detected up to 100 keV.  }
\end{center}
\label{tab_fluxes}
}
\end{table*}
}

\section{Results: the spectral components}\label{results}

\rotate{
\begin{figure*}
\begin{center}
\includegraphics[angle=0,scale=0.38]{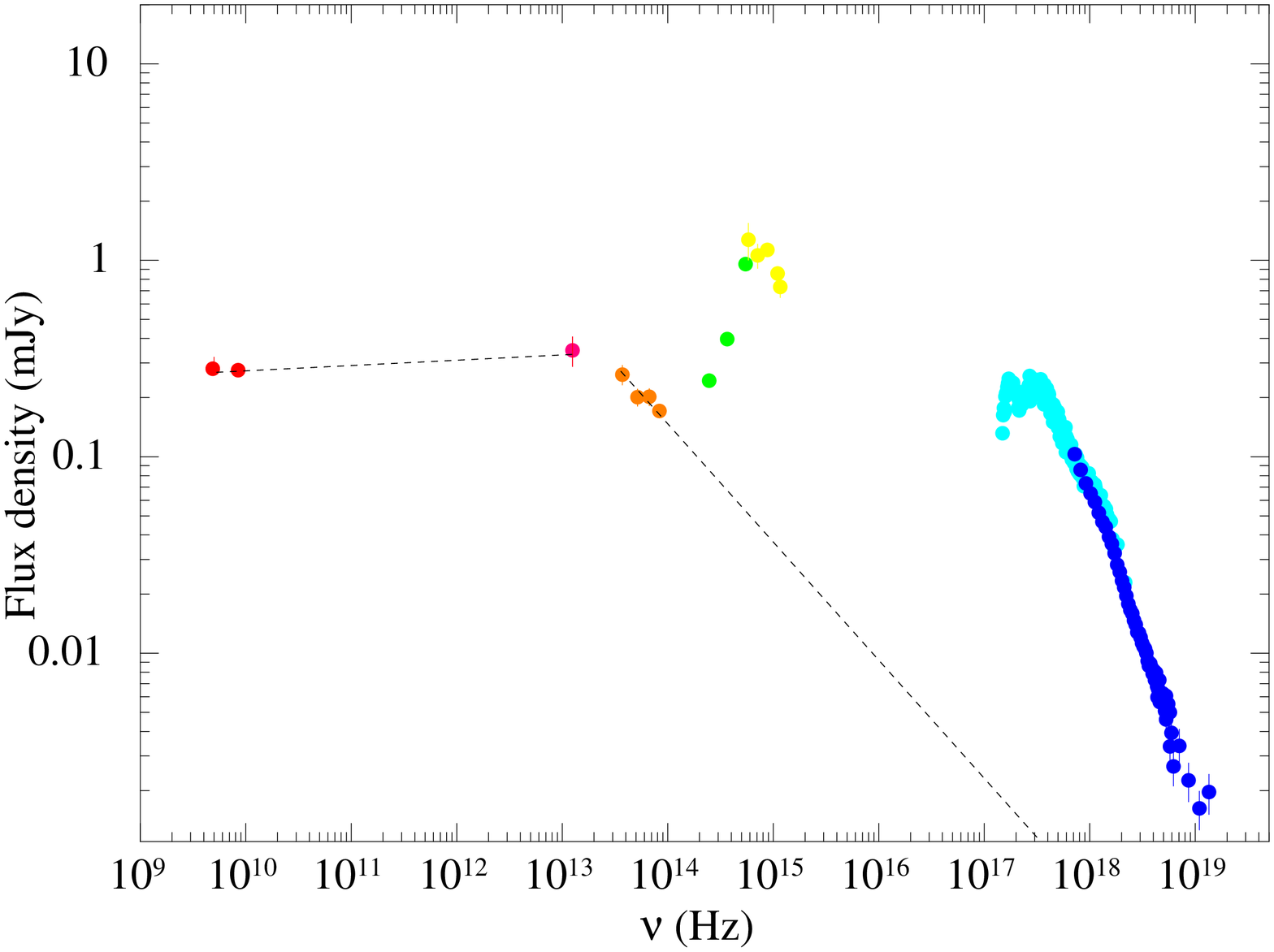}\includegraphics[angle=0,scale=0.38]{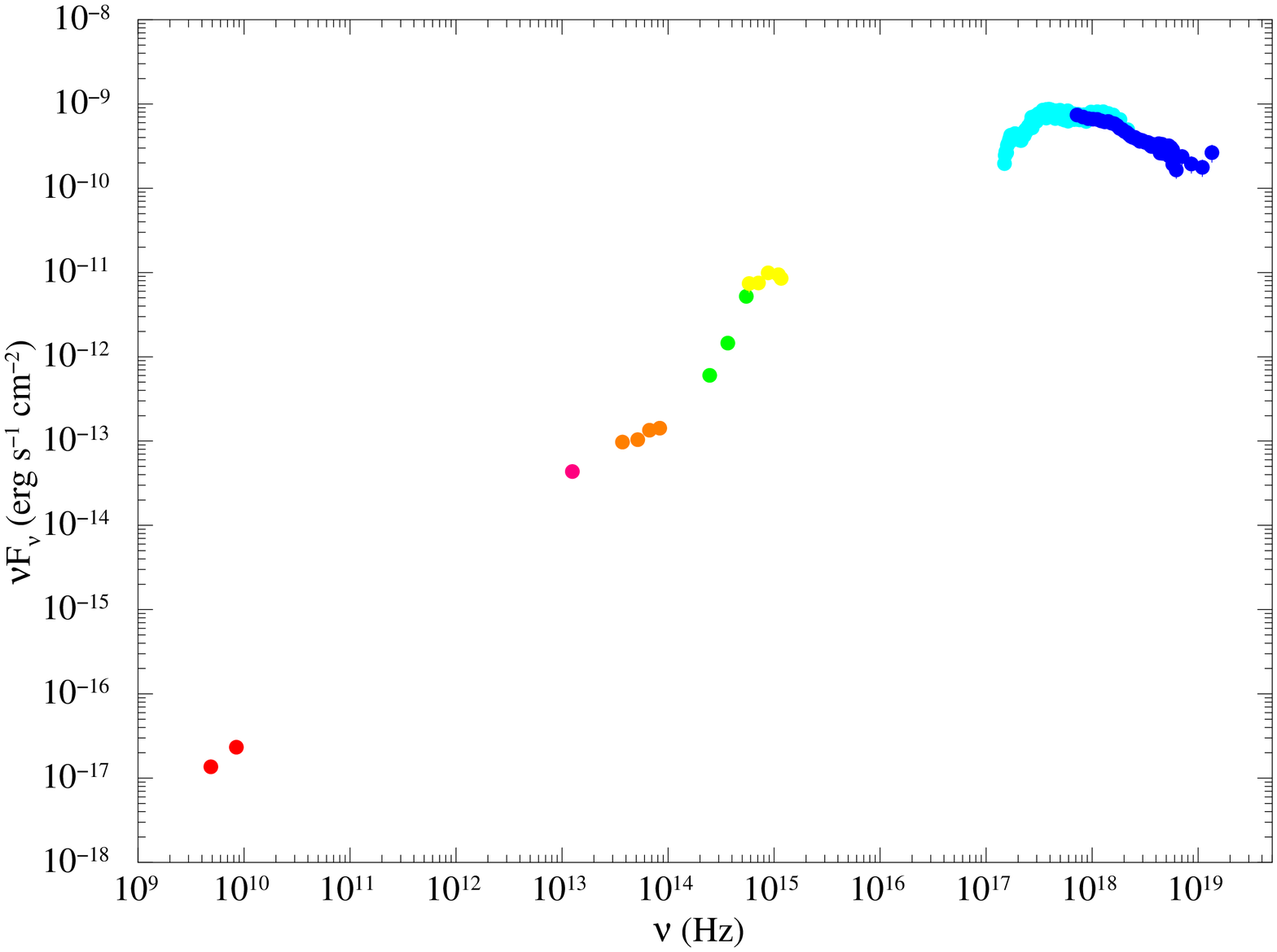}
\caption{Left: Multi-wavelength  observations of the NS X-ray binary 4U~0614+091: VLA (red) Spitzer-MIPS (magenta), Spitzer-IRAC (orange), SMARTS (green), Swift-UVOT (yellow), Swift-XRT (light blue) and RXTE (blue), showing the flat optically-thick and the break to the optically-thin synchrotron emission from the `compact' jet (dashed lines). The optical and IR fluxes are already de-reddened. See \S~\ref{results}. Right:  $\nu$F$_\nu$ representation of the left panel. }
\label{sed}
\end{center}
\end{figure*}
}

In Fig.~\ref{sed} we show the spectral energy distribution of 4U~0614+091. We detected for the first time the radio counterpart at the coordinates: RA $06^{h}17^{m}07^{s}.348\pm0^{s}.007$, Dec $09^{\circ}08\arcmin13\arcsec.44\pm0\arcsec.11$, consistent with the optical position of 4U~0614+091 in the USNO B1.0 catalog: RA $06^{h}17^{m}07^{s}.35\pm0^{s}.003$,   
Dec $+09^{\circ}08\arcmin13\arcsec.6\pm0\arcsec.09$. 

Using our new VLA  radio position in 2006 and the optical position of the USNO B1.0 catalog in 1978, we derive a proper motion upper limit of  $30\pm115$~mas in RA and $160\pm140$~mas in Dec, while the expected LSR offset is of -13.4~mas in RA and 4.2~mas in Dec.  
The average proper motion of isolated pulsars due to the supernovae kick is 200-500~km/s, which would correspond to $\sim300-800$ mas at the distance of 4U~0614+091, over the 28 years between 1978 and 2006 (e.g., Lyne \& Lorimer 1994; Lorimer et al. 1997; Cordes \& Chernoff  1998). Therefore, our upper limits indicate that 4U~0614+091 received a relatively small kick, below-average compared to isolated pulsars. 

Fitting the radio source as a point source, we measure a peak flux density of $0.281\pm 0.010$~mJy/beam at 4.86~GHz and $0.276\pm0.014$~mJy at 8.46~GHz. These fluxes give a radio spectral index of $\alpha=-0.033\pm0.109$.

Analyzing the two observations on November 1 and 2 separately, we measure a flux density, respectively at 8.46~GHz and 4.86~GHz, of $0.33\pm0.01$~mJy and $0.34 \pm0.02$~mJy on November 1 and  of $0.13\pm0.02$~mJy and $0.08\pm0.02$ on November 2. This indicates variability on timescales of less than one day. Moreover, the source was further detected at 4.86~GHz on March 27, 2008 with a flux density of $0.18$~mJy (see \S~\ref{discussion_imaging}). and on May 31, 2009 with a flux density of $\sim0.20$~mJy. A previous radio observation of 4U~0614+091, taken on April 24, 2001 gave a non detection with a $3\sigma$ upper limit of $\sim0.1$~mJy at 4.86~GHz (Migliari \& Fender 2006). Comparing with the RXTE/ASM light curve of 4U~0614+091, we note that when the source was not detected in the radio band on April 24, 2001, 4U~0614+091 was in an unusual low-flux X-ray state. On the other hand, in all the other observations when the radio counterpart was detected, a higher X-ray flux with rapid weekly flares and consequent fast transition to a softer state are observed (Kuulkers et al. 2009). This is somewhat similar to what has been observed in the NS XRB 4U~1728-34, where the radio emission was stronger when the source was in a higher luminosity state, close to the state transition between the hard (island) and the lower-soft (lower-banana) state (Migliari et al. 2003).  

Spitzer/MIPS detected 4U~0614+091 at $24~\mu$m with a flux density of $0.35\pm0.06$~mJy (see Table~1). 
The radio-to-mid-IR spectral index is consistent with being flat with a spectral index $\alpha =0.03\pm0.04$, indicating an optically thick synchrotron spectrum up to the mid-IR (Fig.~\ref{sed}, left). 

Spitzer/IRAC detected 4U~0614+091 with flux densities of $0.26\pm0.02$~mJy at $3.6\mu$m, $0.20\pm0.02$~mJy at $4.5\mu$m, $0.20\pm0.02$~mJy at $5.8\mu$m and $0.17\pm0.01$~mJy at $8\mu$m and a spectral index of $\alpha=-0.47\pm0.15$ (see Table~1). Fluxes and spectral indexes are consistent with the previous IRAC detections of the source on October 25, 2005 (Migliari et al. 2006).  
The quasi-simultaneous observations with MIPS and IRAC allowed us, for the first time, to detect the spectral break between the optically-thick and optically-thin portions of the synchrotron spectrum in a compact jet from a NS XRB, for which we only had upper limits. (Note that the soft X-ray emission of the source remains steady between the IRAC and the MIPS observations.) The break frequency is constrained to be  $1.25\times10^{13}$~Hz $<\nu_{break}< 3.71\times10^{13}$~Hz.

The near-IR data taken with SMARTS show the Rayleigh-Jeans tail of a thermal black-body like emission (as already observed with non-simultaneous data in Migliari et al. 2006). We note that the Swift/UVOT fluxes in the optical/UV show a shape that resembles the spectrum of a single temperature black body from a star or that of an irradiated disk with an optical hump (e.g. Hynes et al. 2002). However, given the faint white dwarf or He-rich partially degenerated companion (e.g. Nelemans et al. 2004, Shahbaz et al. 2008a), we expect the disk emission to dominate over the star. Furthermore, with the small disk of this ultra-compact XRB, we also expect the hump due to irradiation of the disk to be more towards the far-UV. This unexpected spectral shape might be due to the uncertainties in the de-reddening, which become larger and significant in the UV (see \S~\ref{data_swift}). 

\begin{figure}
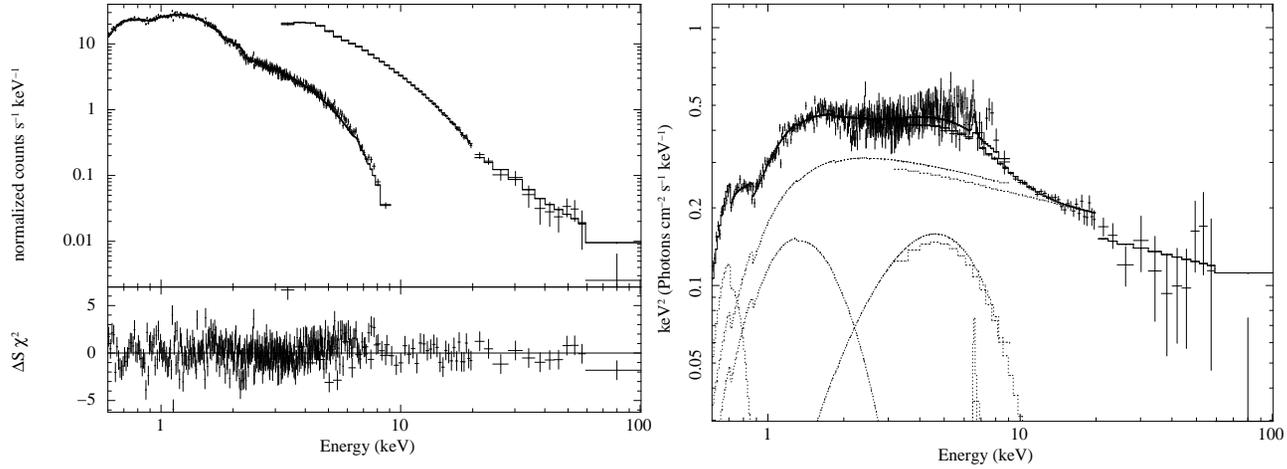

\includegraphics[angle=-90,scale=0.35]{./figure/figure2.ps}\includegraphics[angle=-90,scale=0.35]{./figure/figure2b.ps}
\caption{Left: Swift (XRT) and RXTE (PCA and HEXTE) energy spectrum of 4U~0614+091 in the range 0.6-100 keV. The fitting model consists of two black bodies, an iron Gaussian emission line and a power law, corrected for photoelectric absorption. A Gaussian emission line around $0.62$~keV, possibly an oxygen line, is also needed (see Piraino et al. 1999; Juett et al. 2001). See Table~2. Right: Unfolded energy spectrum showing the spectral components: Gaussian with centroid energy fixed at 0.62 keV, disk black body with kT$_{in}\sim0.35$~keV, black body with kT$\sim1.16$, Gaussian with centroid energy $\sim6.6$~keV, power law with photon index $\sim2.2$. See Table~2. }
\label{xray_spec}
\end{figure}


The continuum of the X-ray spectrum has been modeled in the range 0.6-100~keV using multi-temperature disk black body and a black body with  temperatures of $\sim0.3$~keV and $\sim1.2$~keV, respectively interpreted as the emission from the disk and the surface of the NS, plus a power law with spectral index $\alpha\sim-1.3$, to account for  the non-thermal X-ray emission which is observed without any evidence for a cutoff up to 100~keV.  
Two Gaussian emission lines are also needed: one around $0.62$~keV, possibly an oxygen line (see Piraino et al. 1999; Juett et al. 2001), and an iron line around $6.6$~keV. 
The 2-10 keV unabsorbed flux is $1.0\times10^{-9}$~erg s$^{-1}$ cm$^{-2}$ and the total 0.6-100 keV flux is $3.1\times10^{-9}$~erg s$^{-1}$ cm$^{-2}$. Note that the X-ray flux is comparable to that observed on October 25, 2005, when we first detected 4U~0614+091 with IRAC (Migliari et al. 2006), and indeed the fluxes in the IRAC bands are also comparable.

\begin{table}[]
\begin{center}
\caption{Best-fit parameters of the 0.6-100 keV spectrum of 4U~0614+091 observed with XRT (Swift) and PCA and HEXTE (RXTE). The spectrum is well fit with a multi-color disk black body (dk), a black body (BB), two Gaussian emission lines (ga) and a power law (PL) corrected for photoelectric absorption. The normalization of the power law ($N_{PL}$) at 1 keV is in units of photons keV$^{-1}$ s$^{-1}$. The normalization of the black body (N$_{BB}$) is in units of R$^{2}_{km}$/ D$^{2}_{10}$, where R$_{km}$ is the source radius in km and D$_{10}$ is the distance to the source in units of 10 kpc. The errors quoted are at the 90\% confidence level. The normalization of the Gaussian lines N$_{ga}$ is in units of total photons $cm^{-2} s^{-1}$ in the line}

\begin{tabular}{cc}

\hline
\hline
N$_{\rm H}$ ($\times 10^{22}$~cm$^{-2}$)& $0.34\pm0.03$ \\
\hline
E$_{ga1}$ (keV) & 0.62(fixed) \\
$\sigma_{ga1}$ (keV) & $0.103\pm0.003$ \\
N$_{ga1}$ & $0.38\pm0.06$ \\
\hline
E$_{ga2}$ (keV) & $6.6\pm0.2$ \\
$\sigma_{ga2}$ (keV) & $<0.7$ \\
N$_{ga2}$ $(\times 10^{-4})$& $4.5_{-1.3}^{+3.4}$ \\
\hline
kT$_{dk}$ (keV)&  $0.35\pm0.02$\\
N$_{dk}$ $(\times 10^{3})$& $2.4\pm0.4$ \\
\hline
kT$_{BB}$ (keV)& $1.16\pm0.02$ \\
N$_{BB}$ & $16.7\pm1.0$ \\
\hline
$\Gamma_{PL}$ & $2.24\pm0.02$ \\
N$_{PL}$ & $712\pm20$ \\
\hline
$\chi^{2}_{red}$ (d.o.f.) & 1.81 (440) \\
\hline
\end{tabular}
\end{center}
\label{tab_fit}
\end{table}

\section{Discussion}\label{discussion}
\subsection{Jet spectrum, break frequency and jet power}\label{discussion_jetspec}

The low-frequency spectrum of 4U~0614+091 (4.86~GHz to 8~$\mu$m) is consistent with synchrotron emission from a continuously replenished compact jet, as described in its first formulation by Blandford \& K\"onigl (1979). The spectrum is consistent with being optically thick from the radio to the mid-IR band, with a flat spectrum of spectral index $\alpha=0.03\pm0.04$.  
 At a frequency $1.25\times10^{13}$~Hz$<\nu_{break}<3.71\times10^{13}$~Hz a break occurs and the spectrum steepens to a power law with spectral index $\alpha = -0.47\pm0.15$ at least up to $8\mu$m (i.e., $8.3\times10^{13}$~Hz). In the near-IR band the accretion disk starts to dominate the emitting spectrum. 

The synchrotron break frequency for the NS 4U~0614+091 is lower than the measured break frequency in the BH GX~339-4 by a factor of $\sim10$ (Corbel \& Fender 2002; see Migliari et al. 2006).  The lack of statistics and the fact that the break frequency has been observed to vary within the same source (in GX~339-4), possibly in relation to variations of the bolometric luminosity, makes it premature to generalize any possible relation between the $\nu_{break}$ and the nature of the compact object.  However, under certain assumptions, we can make some considerations on their possible dependency, as follows. 
 
The break frequency is the frequency above which the jet is fully transparent to its emitting synchrotron radiation. The association between this break frequency and the geometrical parameters of the jet is still unknown. 
If we relate an emitting frequency to a specific position in the jet, parametrized by its distance from the jet base, and make the disk-jet symbiosis assumption that a fixed fraction of the mass accretion rate is channeled into the jet (Falcke \& Biermann 1995), we may have some clues on the jet geometry. In the latest prescription of the disk-jet symbiosis model by Falcke et al. (2004), the  $\nu_{break}$ is related to the size of the base of the jet $R_{nozzle}$ and to the mass accretion rate $\dot{M}$, by $\nu_{break} \propto \dot{M}^{2/3} R_{nozzle}^{-1}$.

Corbel \& Fender (2002) detected the $\nu_{break}$ in two observations of the BH GX~339-4 during its hard state (in one of the two, only a lower limit). The observation with the lower bolometric luminosity shows the higher $\nu_{break}$.
Therefore, assuming that at a higher bolometric luminosity corresponds a higher mass accretion rate, the predicted positive dependence of the $\nu_{break}$ to the $\dot{M}$ seems to be in contrast with the observations of the BH GX~339-4. 
Moreover, in the prescription of Falcke et al. (2004) the jet is attached to the inner radius of the disk. 
i) If the disk is truncated, a lower luminosity would correspond to a lower mass accretion rate and a higher $R_{nozzle}$, implying a lower $\nu_{break}$. ii) If instead the inner radius corresponds to the innermost stable orbit also in the hard state, $R_{nozzle}$ should not change significantly with the accretion rate, and a lower bolometric luminosity would correspond to a lower mass accretion rate and therefore, according to the above formula, to a lower $\nu_{break}$. 

In the NS 4U~0614+091, the X-ray bolometric luminosity, i.e. in the range 0.6-100 keV\footnote{Given the spectral shape of the X-ray spectrum, with a blackbody peaking at $\sim1$~keV and a power law with photon index $>2$, the X-ray bolometric correction is very small, of the order unity.}, is $3.4\times10^{36}$ erg s$^{-1}$ (at 3 kpc) and the break frequency is $\sim1-4\times10^{13}$~Hz. In the BH GX~339-4 observation where the break frequency was detected, the X-ray bolometric luminosity was $4\times10^{37}$ erg s$^{-1}$ (1-200~keV integrated luminosity, using a distance of 4~kpc: Corbel \& Fender 2002) and the break frequency $\sim2\times10^{14}$~Hz. 
Comparing these two observations, both the total accretion power and $\nu_{break}$ of GX~339-4 are higher than those of 4U~0614+091. There is an anticorrelation between the accretion power (i.e. X-ray bolometric luminosity) and the break frequency in the two GX~339-4 observations, therefore, if the accretion power--$\nu_{break}$ variations do not depend on the nature of the compact object, we would expect 4U~0614+091 to have a higher break frequency than GX~339-4. 
The lower break frequency observed in 4U~0614+091 would therefore point towards either a different jet formation mechanism between BHs and NSs or a different geometry of the jet (smaller in the NS) or less power channeled into the jet (i.e. the disk-jet symbiosis hypothesis is not valid), maybe due to an effect of the specific compact object, such as the presence of the magnetic field in the NS.

Given the X-ray bolometric luminosity of  $\sim3.4\times10^{36}$ erg s$^{-1}$ and using $\alpha = -0.47$, the integral of the optically thin part of the synchrotron spectrum, from the thin-thick break frequency, is $\sim5.7 \times 10^{32}\times$E$_{c}^{0.53}$ erg/s, where E$_{c}$ is the high-energy cutoff of the synchrotron spectrum in eV. The optically thin synchrotron emission is observed up to 0.3~eV, which corresponds to a lower limit on the radiative power of the jet of $3\times10^{32}$erg s$^{-1}$. 

The spectrum leads to a jet power of  $\sim1.1 \times10^{34} \mu_{0.05}^{-1} $E$_{c}^{0.53}$ erg s$^{-1}$, where $\mu_{0.05}$ is the radiative efficiency of the synchrotron emission in units of 0.05 [0.05 is the upper limit estimated for GRS~1915+105 by Fender \& Pooley (2000), consistent with what is found in AGN  (i.e. $\mu<0.15$; Celotti \& Ghisellini 2003), and the theoretical prediction of $\mu<0.15$ in the jet model of Blandford \& K\"onigl (1979)]. The jet power to X-ray bolometric luminosity ratio is  L$_{J}$/L$_{Xbol}$$\sim2\times10^{-3}\times \mu_{0.05}^{-1} $E$_c^{0.53}$.

\subsection{Imaging the compact Jet}\label{discussion_imaging}

Brightness temperature arguments, polarization studies (Shahbaz et al. 2008b; Russell et al. 2008) and spectral studies indicate that the out-flowing matter in NSs during a steady state is in the form of a self-absorbed conical jet.
In BHs this evidence have been corroborated by observations of spatially resolved jets of two systems (Stirling et al. 2001; Dhawan et al. 2000). However, a compact jet has never been spatially resolved in a NS.
We were awarded 2.5 h of time with the HSA (VLBA+Arecibo+GBT+phased VLA) to attempt to detect the compact jet in 4U~0614+091. The source was observed on March 27, 2008 at 4.86~GHz. The standalone data from the phased VLA measured a source with flux density $\sim0.18$~mJy, 30\% less than the flux density measured during our multiwavelength campaign (Table~1); this is indicative of radio variability.
However, the short duration of the observations, together with the extremely high sensitivity of the three big dishes (the GBT, phased VLA, and  especially Arecibo) meant that the side-lobes of the naturally-weighted dirty beam were too high (91\% of the peak) to identify  accurately the source (with a nominal rms of $23\mu$Jy), particularly given the lack of an accurate position from other wavelengths. The VLA was in C configuration at the time, giving a position accurate to 200 mas, with respect to an HSA  beamsize of $3.7\times0.8$~mas$^{2}$. A compact jet as inferred from our spectral study (this work) has never been spatially resolved. 

\subsection{Non-thermal Hard X-ray vs. Synchrotron-Self Compton from the jet}\label{discussion_hardtail}

Hard X-ray tails in X-ray binaries are well fitted with a model where thermal accretion disk seed photons are Compton up-scattered by a `corona' of hot plasma. Markoff, Nowak \& Wilms (2005) analyzed a sample of BH XRBs' energy spectra and showed that fits of the hard X-ray tails using jet model's components (i.e. synchrotron and SSC) are statistically as good as single-component Comptonizing corona models (i.e. eqpair; Coppi 1999). This result would be in agreement with the possibility that the `corona' of hot plasma might be in fact the base of the jet itself, at least in BHs.  
In NS X-ray binaries the hard X-ray tails can also be well fitted by corona Comptonizing models with thermal electron with temperatures of 25-30~keV for low-luminosity NSs (e.g. Barret et al. 2000) tens of keV for accreting millisecond X-ray pulsars (e.g. Gierlinski, Done \& Barret 2002; Poutanen \& Gierlinski 2003; Gierlinski \& Poutanen 2005; Falanga et al. 2005a, Falanga  et al. 2005b, Falanga et al. 2007) and a few keV for brighter sources (e.g. Paizis et al. 2006). 
Fits of energy spectra using the broadband jet model have not been attempted yet for NSs. In this work, we have detected both the optically-thin spectrum of the jet and the hard X-ray tail in 4U~0614+091. Therefore, we are able for the first time to explore the possibility that in NSs, like BHs, the high-energy non-thermal energy spectrum might be reproduced, at least partially, with radiative processes taking place in the jet.

From the broadband spectrum in Fig.~\ref{sed}, left, we see that the synchrotron emission alone, estimated as the extrapolation  of the optically-thin jet spectrum observed, at high energies (dashed line), cannot reproduce the high-energy X-ray tail observed (as it does in e.g. the BH GX~339-4; Corbel \& Fender 2002).  As a first step, we test whether the hard X-ray tail could be produced by SSC emission from the same population which produces the synchrotron emission (i.e. post-shock accelerated jet;  Markoff, Nowak \& Wilms 2005). 
We therefore constructed a simple, robust one-zone model for both synchrotron and SSC emission (e.g., Kirk, Rieger, \& Mastichiadis 1998).
The emission region is taken to be a cylinder with variable aspect ratio and size, with the same population of electrons being responsible for synchrotron and inverse Compton emission.  This simple model is appropriate, given that both the observed optically thin synchrotron tail and the SSC emission should come from the base of the jet (where photon and particle densities are highest). Different geometries (e.g., spherical) change the results only by factors of order unity.

Assuming uniform plasma density and isotropic synchrotron emission in the frame of the plasma, the SSC emission is then calculated analytically (e.g., Rybicki \& Lightman 1979).

Since the emission is unresolved the problem is underconstrained and we have several free parameters at our disposal that can be varied in order to find a suitable SSC solution that is consistent with the observed spectrum:
\begin{itemize}
\item{Radius and length of the cylindrical emission region.}
\item{The bulk Lorentz factor of the emitting plasma.}
\item{The viewing angle of the jet (relative to the direction of motion)}
\item{The lower cutoff frequency of the synchrotron spectrum at the   base of the jet (which is unobservable due to self-absorption).}
\end{itemize}
The remaining parameters are determined by the observed synchrotron luminosity and spectrum.  We then attempt to match the observed hard tail by varying the free model parameters.

We find that matching the observed high X-ray flux is not feasible for reasonable physical parameters: for reasonable jet bulk Lorentz factors of order $\Gamma \lesssim 5$, the observed flux would require unrealistically low magnetic field values, roughly $\beta \sim 10^4$ (that is, the field is four orders of magnitude out of equipartition). For reasonable magnetic field values, with a $\beta \lesssim 10$, the required Lorentz factors would require a bulk Lorentz factor of order 50 or larger, with a viewing angle close to $90^{\circ}$.  Roughly, the relationship between the equipartition fraction $\beta$ and $\Gamma$ is $\beta \sim 10^{-4}(\Gamma/5)^{2.5}$.

However, even if the jet is either extremely undermagnetized or extremely relativistic, the spectral slope of the predicted SSC emission is too hard to match the observed X-ray spectrum: The optically thin synchrotron spectrum, which is imprinted on the SSC spectral slope, is -0.47, while the observed hard tail has a spectral slope of -1.25.  We conclude that SSC emission alone, from the post-shock jet particles, is {\it not} a viable mechanism to explain the observed hard X-ray tail of NS 4U~0614+091. 
This result indicates that an extra Comptonizing population of hot plasma apart from that producing the direct synchrotron emission in the jet, exists close to the NS in the form of a corona where the hard X-ray emission is associated with the accretion disk boundary layers (e.g. Mitzuda et al. 1984;  Klu\'zniak \& Wilson 1991),  a hot inner flow, an ADAF (Narayan et al. 1996;  Barret et al. 2000; Medvedev \& Narayan 2001; see also the discussion in Done \& Gierlinsky 2003), a mildly-outflowing corona (e.g. Beloborodov 1999), or a pre-shock, quasi-thermal, component at the base of the jet (Markoff \& Nowak 2004). 
To assess in particular this last possibility,  a broadband fit of the spectrum with a disk-jet model is necessary.

\section{Conclusions}

We present here the most complete energy spectrum of the NS XRB 4U~0614+091 to date, from the radio to the hard X-ray band. 

{\it 1)} We detected for the first time the radio counterpart of 4U~0614+091 with flux densities of $\sim0.3$~mJy, at both 5 GHz and 8.5 GHz (\S~\ref{results}). The coordinates are:  RA $06^{h}17^{m}07^{s}.348\pm0^{s}.007$, Dec $09^{\circ}08\arcmin13\arcsec.44\pm0\arcsec.11$

{\it 2)} We detected for the first time the source at $24~\mu$m and obtained the first measurement of the actual `break frequency' of the compact jet spectrum in a NS XRB, which is $\sim(1-4)\times10^{13}$~Hz (\S~\ref{results}). 

{\it 3)} We measured the spectral index of the optically thick part of the synchrotron spectrum, which is consistent with being flat from the radio to the mid-IR band (\S~\ref{results}). 

{\it 4)} We confirmed the existence of an optically thin part of the jet, which is evident in the IR band and can be well fitted by a power law with spectral index $\sim-0.5$  (\S~\ref{results}).

{\it 5)} There is an anti-correlation between the accretion power (i.e. X-ray bolometric luminosity) and the break frequency in the two  observations of the BH GX~339-4 where the synchrotron IR break frequency has been observed. This seems to be in contrast with the prescription: $\nu_{break} \propto \dot{M}^{2/3} R_{nozzle}^{-1}$, in the disk-jet symbiosis hypothesis. 
Moreover, in 4U~0614+091 both the accretion power and the synchrotron break frequency are lower than that in the BH. This is in contrast with the trend observed in GX~339-4, pointing towards either a different jet formation mechanism between BHs and NSs or a different geometry of the jet (smaller in the NS) maybe due to an effect of the specific compact object on the jet power, such as the presence of the magnetic field in the NS (\S~\ref{discussion_jetspec}). 

{\it 6)} The lower limit on the radiative jet power is  $\sim3\times10^{32}$~erg s$^{-1}$ (i.e. observed up to 0.3 eV) and the total jet power is $\sim1.1 \times10^{34} \mu_{0.05}^{-1} $E$_{c}^{0.53}$~erg/s  (where E$_{c}$ is the high-energy cutoff of the synchrotron spectrum in eV and $\mu_{0.05}$ is the radiative efficiency in units of 0.05). The lower limit on the fraction of the accreting bolometric power L$_{Xbol}$ channeled into the jet is L$_{J}$/L$_{Xbol}$$\sim2\times10^{-3}\times \mu_{0.05}^{-1} $E$_c^{0.53}$.

{\it 7)}  The detection of the optically thin part of the jet together with the X-ray tail above 30 keV, allowed us to assess  the possible jet contribution to the non-thermal hard X-ray emission via synchrotron self-Compton processes. We conclude that post-shock SSC emission alone is {\it not} a viable mechanism to explain the observed hard X-ray tail of 4U~0614+091. This result points towards  the existence of an extra Comptonizing region as an ADAF corona, or a pre-shock, quasi-thermal component at the base of the jet (\S~\ref{discussion_hardtail}). 

{\it 8)} We measured no significant variations of the IR with respect to the previous IRAC observation, both taken in 2006. The radio emission varied from $<0.1$~mJy in 2001 to $\sim0.3$~mJy in 2006  and to $\sim0.18$ mJy  in 2008 (\S~\ref{discussion_imaging}). The low radio upper limit in 2001 might be related to a steady, low-luminosity X-ray state. 

{\it 9)} We found the spectral evidence that in 4U~0614+091 a compact jet of the same kind observed in BHs (stellar mass and AGN) exists. This, together with the fact that the system is located at a distance of only $3.2$~kpc, make this source our best candidate in order to try to spatially resolve its jet with high-spatial resolution radio interferometry. We obtained HSA observations of 4U~0614+091, but, possibly for technical issues, we were not even able to identify it (\S~\ref{discussion_imaging}). 

\acknowledgments

SM would like to thank Sera Markoff for uncountable discussions and Erik Kuulkers also for sharing the results of his work before publication.  
SM is an ESA Fellow. JMJ is a Jansky Fellow. EG is a Hubble Fellow, supported through Postdoctoral Fellowship grant number HST-HF-01218.01-A from the Space Telescope Science Institute, which is operated by AURA Inc., under NASA contract NAS5-26555.. The National Radio Astronomy Observatory is a facility of the National Science Foundation operated under cooperative agreement by Associated Universities, Inc. This work is based on observations made with the {\it Spitzer} Space Telescope, which is operated by the Jet Propulsion Laboratory, California Institute of Technology under a contract with NASA. Support for this work was provided by NASA through an award issued by JPL/Caltech. SM was partially supported by Spitzer grant 1289781. JAT was partially supported by Spitzer grant 1321162.

\end{document}